\documentclass[useAMS,usenatbib]{mnras}
\usepackage{float}
\input{boxed1.sty}
\floatstyle{simplerule}
\restylefloat{figure}
\usepackage{amssymb}
\usepackage{graphicx}
\usepackage[table]{xcolor}

\definecolor{grey80}{rgb}{0.90,0.90,0.90}

\usepackage{amsmath}

\newcounter{mycount}

\title[The Rafita asteroid family]{\centering \bf The Rafita asteroid family}
\author[S. Aljbaae, V. Carruba, J. R. Masiero, R. C. Domingos, and M. Huaman]
{S. Aljbaae$^{1}$\thanks{E-mail: \href{safwan.aljbaae@feg.unesp.br}{safwan.aljbaae@feg.unesp.br}}, V. Carruba$^{1}$, J. R. Masiero$^{2}$, R. C. Domingos$^{3}$ and M. Huaman$^{1}$\\
  $^{1}$Univ. Estadual Paulista (UNESP),  Faculdade de Engenharia, Guaratinguet\'a\,, CEP 12516-410, SP, Brazil\\
  $^{2}$Jet Propulsion Laboratory/Caltech, 4800 Oak Grove Dr., MS 183-601, Pasadena, CA 91109, USA.\\
$^{3}$Univ. Estadual Paulista (UNESP), S\~{a}o Jo\~{a}o da Boa Vista, SP, CEP 13874-149, SP, Brazil.}

\begin{document}

\date{Accepted 2017 January 19. Received 2017 January 17; in original form 2016 September 5.}

\pagerange{\pageref{firstpage}--\pageref{lastpage}} \pubyear{2016}

\maketitle
\label{firstpage}
% ============================
\begin{abstract}

  The Rafita asteroid family is an S-type group located in the middle main belt,
  on the right side of the 3J:-1A mean-motion resonance.  The proximity of this
  resonance to the family left side in semi-major axis 
  caused many former family members to be lost. As a consequence, the family
  shape in the $(a,1/D)$ domain is quite asymmetrical, with a preponderance
  of objects on the right side of the distribution. The Rafita family is also
  characterized by a leptokurtic distribution in inclination, which allows the
  use of methods of family age estimation recently introduced for
  other leptokurtic families such as Astrid, Hansa, Gallia, and Barcelona. In
  this work we propose a new method based on the behavior of an asymmetry
  coefficient function of the distribution in the $(a,1/D)$ plane to date
  incomplete asteroid families such as Rafita. By monitoring the time behavior
  of this coefficient for asteroids simulating the initial conditions at the
  time of the family formation, we were able to estimate that the Rafita
  family should have an age of $490\pm200$ Myr, in good
  agreement with results from independent methods such as Monte Carlo
  simulations of Yarkovsky and Yorp dynamical induced evolution and the
  time behaviour of the kurtosis of the $\sin{(i)}$ distribution.
   Asteroids from the Rafita family can reach orbits similar to 8\% of
    the currently known near Earth objects. $\simeq$1\% of
    the simulated objects are present in NEO-space during the final 10 Myr
    of the simulation, and thus would be comparable to objects in the
    present-day NEO population.

\end{abstract}
% ============================
% ============================
\section{Introduction}
\label{Introduction}

Asteroid families are groups of minor planets originated by catastrophic
disruption events of single parent bodies, identified by clustering in
their proper orbital elements, which are very close to invariants of
motion characterizing their orbits.  Typical asteroid family shows a clear
V-shape in the distribution of associated asteroids in the
(proper $a$, $H$) or (proper $a$, 1/$D$) plane, where $a, H$, and $D$ are the
asteroid proper semi-major axis, absolute magnitude and diameter,
respectively. \citet{morbidelli_1995} showed that families located near
the boundary of resonant zones in the main asteroid belt might have
importantly contributed to the near Earth material and, consequently,
also to the cratering history of the Earth or Moon. For instance,
asteroids near the 3J:-1A resonance with eccentricities less than 0.1
would undergo small variations in eccentricity on timescales of million years
and then suddenly they undergo eccentricity increases to over 0.3, becoming
a Mars-crossers, and a possible source for Earth-crossing
materials \citep{wisdom_1983}.

Among main belt families, the Rafita family is an S-complex group.
1295 members of this family were identified in \citet{nesvorny_2015}
using the Hierarchical Clustering Method (HCM), described in
\citet{bendjoya_2002}, with a distance cutoff of $70 ~ m/s$.
The Rafita family is located in the central main belt between the
3J:-1A and 11J:-4A mean-motion resonances. Its values of the
$a$, $e$, $\sin(i)$ proper orbital elements go from 2.53 to 2.65 au,
from 0.15 to 0.20, and from 0.11 to 0.15, respectively.
The global structure of this family seems to have been significantly
affected by the 3J:-1A resonance. The family is characterized by only
one side of its V-shape, with its left side presumably missing because of
the interaction of former family members with this resonance \citep{spoto_2015}.

Based on this last hypothesis, we aim in this work to investigate
in detail the local dynamical evolution of asteroids in the Rafita family,
analyzing the past leakage of the family members to the 3J:-1A resonance, and
to check what information on the family age and dynamical evolution of the
peculiar orbital configuration of this group may provide.  In particular, by
quantifying the current level of asymmetry in semi-major axis and 1/$D$ of the
family through the use of a $C$-target function and a newly introduced
asymmetry coefficient $A_S$, and by studying the time behavior of this
parameter for fictitious Rafita family members simulating the initial
conditions of the family at the time of the break-up, we propose to
introduce a new method to obtain age estimates for asteroid groups
interacting with powerful mean-motion resonances.
Preliminary estimates of the contribution to the NEA population
  from the Rafita family will also be attempted in this work.

% ============================
\section{Family identification and physical properties}
\label{sec: fam_ide}

\begin{figure}[!htp]
   \centering\includegraphics[width=0.8\linewidth]{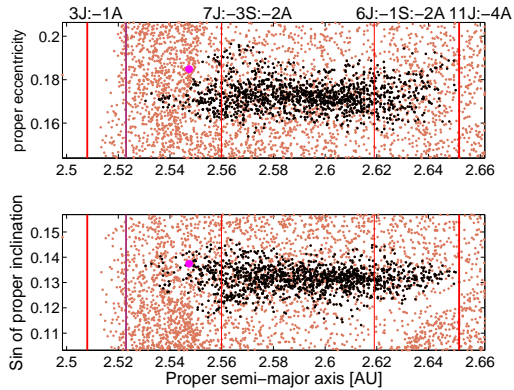}
   \caption{A projection in the ($a$, $e$) (top panel) and ($a$, $\sin (i)$)
     (bottom panel) domains, of members of the HCM Rafita cluster (1295
     members, black dots), and of the local background (3987 members, orange 
     dots). Vertical red lines display the local mean-motion resonances.}
     \label{Fig: Orbital_location}
\end{figure}

A set of 406253 proper elements of numbered asteroids from the
\href{http://hamilton.dm.unipi.it/astdys/}{AstDyS}
\footnote{\href{http://hamilton.dm.unipi.it/astdys/}
  {ftp$:$//hamilton.dm.unipi.it/astdys/}. accessed on March 21st, 2016
database was used in this work to identify 3987 asteroids in the background
of the Rafita family, as identified in \citet{nesvorny_2015}. Objects in the
background were selected if they were within the minimum and maximum values,
$\pm 0.01$ of the $a$, $e$, and $\sin (i)$ of observed members of the Rafita
dynamical group.} In the ($a$, $e$) and ($a$, $\sin (i)$) planes, the orbital
locations of the family members are shown as full black dots in 
Fig.~\ref{Fig: Orbital_location}, where local background asteroids are also
drawn as full orange dots. The HCM family represent about 32\% of the
asteroids in its region.  Vertical heavy and light red lines display the
location of the main two- and three-body mean-motion resonances in the region,
respectively. The indigo line displays the chaotic layer near the 3J:–1A
resonance as defined in \citet{morbidelli_2003}. That corresponds to a zone
within 0.015 au from the resonance boundary, where the asteroid population is
depleted because of the presence of several high-order mean-motion
resonances. The position of 1644 Rafita itself is shown as a magenta full dot.

\begin{figure}[!htp]
  \centering\includegraphics[width=0.8\linewidth]{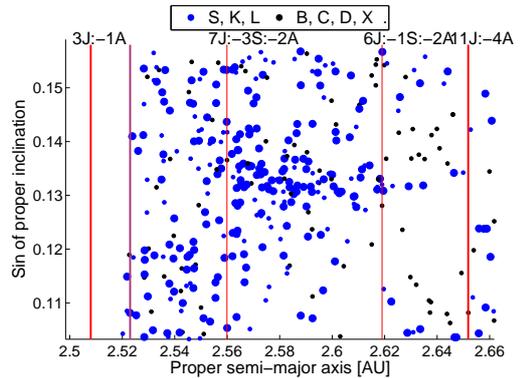}
  \caption{An ($a$,$\sin (i)$) projection of the 374 asteroids in Rafita
    region with taxonomic information from the SDSS-MOC4 database. See the
    figure legend for the meaning of the full dots symbols, other symbols are
    the same as in Fig.~\ref{Fig: Orbital_location}.}
  \label{Fig: Taxonomy}
\end{figure}

We then revised the physical and taxonomic properties of the objects in the
Rafita region. Only nine objects had taxonomic data (4 S-,2 X-, 1 Ch- and 1
Sa- types), observed during the second phase of the Small Mainbelt Asteroid
Spectroscopic Survey (SMASS II\footnote{\href{http://smass.mit.edu/smass.html}{http://smass.mit.edu/smass.html}}), which used the feature-based taxonomy of
\citet{bus_1999} and \citet{bus_2002a, bus_2002b}. Also, using the photometric
data from the fourth Release of the Sloan Digital Sky Survey Moving Object
Catalog (SDSS-MOC4\footnote{\href{http://www.astro.washington.edu/users/ivezic/sdssmoc/sdssmoc.html}
  {ftp$:$//www.astro.washington.edu/users/ivezic/sdssmoc/sdssmoc.html}},
\citep{ivezic_2001}), taxonomic information for 374 asteroids was obtained
with the method described in \citet{demeo_2013}, computing the spectral slopes
over the $g'$, $r'$, and $i'$ reflectance and the $z'-i'$ colours.  We found
191 S-, 55 L-, 45 K-, 38 C-, 34 X-, 8 D-, 2 B- and 1 V-types. Only one
asteroid had taxonomic data in the SMASS II classification in the Rafita HCM
group: 1587 Kahrstedt (SA). 123 asteroids in the HCM group had taxonomic
information in the SDSS-MOC4 data-set: 72 S-, 21 K-, 17 L-, 8 X-, 3 C-, 1 D-,
and 1 V-types. The 12 X-, C-, D- and V-type objects, corresponding to 9.8\%
of the objects with taxonomic information, can be considered as taxonomic
interloper of the S-complex Rafita family, and will be excluded from the
family list. As shown in Fig.~\ref{Fig: Taxonomy}, the S-complex objects
dominate the local background, but with a significant mixing with other
type objects. 

\begin{figure}[!htp]
   \centering\includegraphics[width=0.8\linewidth]{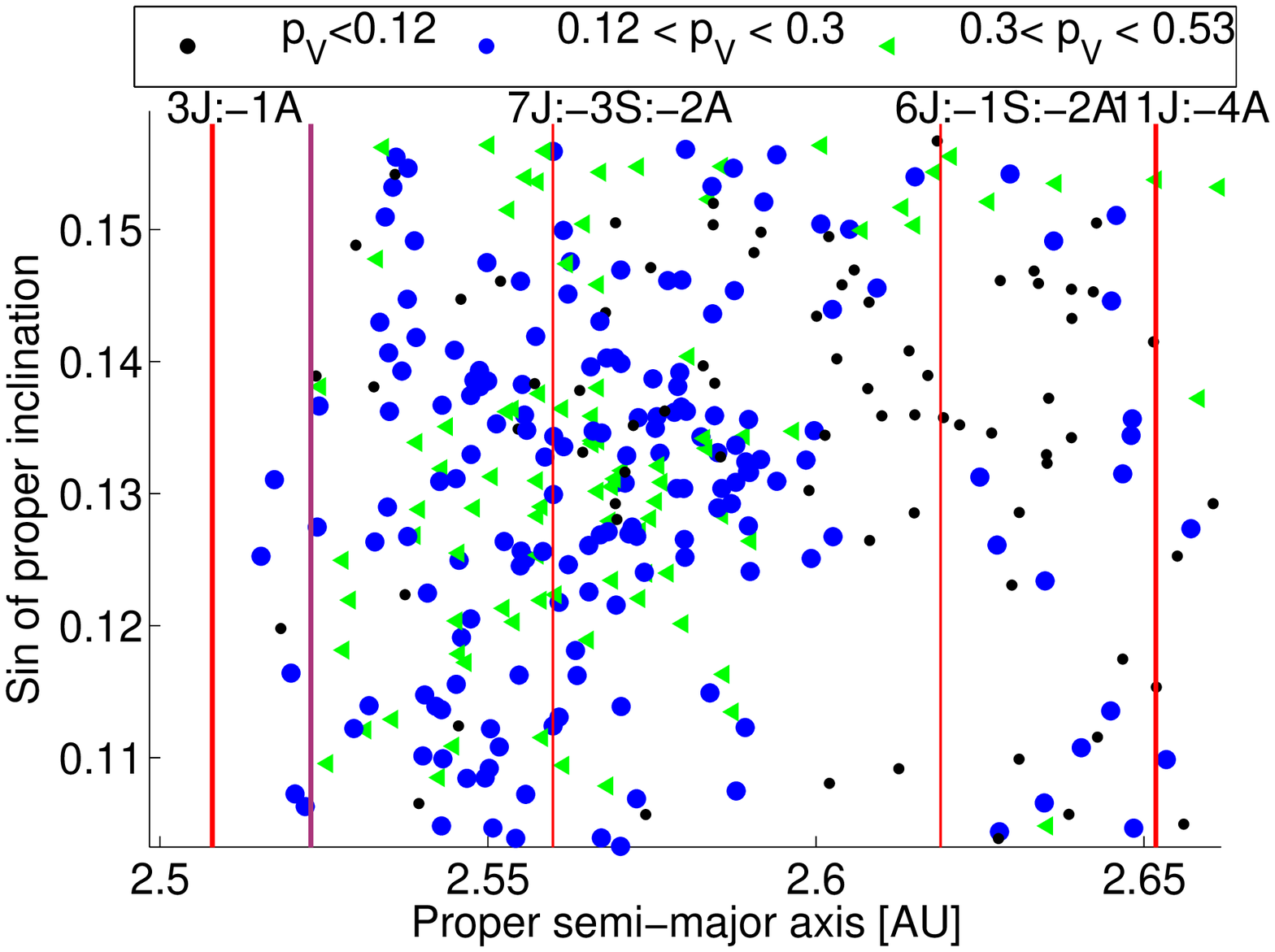}
   \caption{An ($a$,$\sin (i)$) projection of the 449 bodies in Rafita
     region with albedos information from the WISE data-set. See the figure
     legend for the meaning of the full dots symbols, other symbols are the
     same as in Fig.~\ref{Fig: Orbital_location}.}
   \label{Fig03_WISE}
\end{figure}

Regarding the albedo information, the Wide-field Infrared Survey
Explorer (WISE) and Near-Earth Object
WISE~\footnote{\href{sbn.psi.edu/pds/resource/neowisediam.html}
  {sbn.psi.edu/pds/resource/neowisediam.html}} data
\citep{masiero_2011} was used to identify 449 bodies of the local
background with geometric albedos $p_V$ varying between 0.022 and 0.531.
339 of these 449 objects do not have taxonomic information.  About half of
those last albedos values vary between 0.12 and 0.30, and this suggests that
they may be compatible with an S-complex composition.  About 28\% have an
albedo greater than 0.3. The remaining 22\% have low albedos
(lower than 0.12), which is usually associated with a C-complex taxonomy
\citep{masiero_2011}. The ($a$,$\sin (i)$) projection of the 449 objects
with albedos in the region is displayed in the Fig. \ref{Fig03_WISE}.
In the Rafita HCM group, there are 154 asteroids with identified albedos
between 0.036 and 0.53. The mean albedo found in the group is 0.243 and the
median one is 0.247.

An estimate for 154 effective diameters ($D$) of asteroids in the
Rafita HCM group is also provided in the WISE data. The diameters of the
remaining objects were computed according to equation (1) in
\citet{carruba_2003}. The obtained diameters vary between 0.644 km
(323442 2004 HB1) and 17.472 km (6076 Plavec). Therefore, the Rafita family
is formed mostly by small-diameter objects, and could be the result of a
catastrophic disruption event.

The total mass of the family is estimated as $0.618\times 10^{17}$ kg,
assuming a spherical shape, a homogeneous structure for each asteroid
with densities according to the values typical for each class
\citep{demeo_2013}. For objects without taxonomical information,
a mean density value of 2.91 $g~ cm^{-3}$ was assumed. This density is
within the typical densities of S-type asteroids
($2.72 \pm 0.54$, \citep{demeo_2013}). Overall, the total mass of the family
as estimated in this work suggested that the Rafita family originated from
the break-up of $D\simeq 45$ km asteroid in the region, i.e., a medium-sized
asteroid.

% ============================
\section{Dynamical map}
\label{sec: Dynamical_map}

Many features of the asteroid distribution observed in
Fig.~\ref{Fig: Orbital_location} may be explained by the long-lasting
effects of dynamical evolution. Constructing dynamical maps for the region of
Rafita can be useful to improve our understanding of the local dynamics. For
this task, the \textsc{swift\_mvfs} symplectic integrator from the
\textsc{swift} package \citep{levison_1994}, modified by \citet{broz_1999},
was used to integrate 6205 mass-less particles over 20 Myr. The initial
conditions varied between 2.49 and 2.66 au in $a$, and between $1.00^{\circ}$
and $9.6^{\circ}$ in $i$. An equally spaced grid in the ($a$, $\sin (i)$) plane
was generated using 85 intervals in $a$ and 73 in $i$. The initial values of
the eccentricity $e$, longitude of the ascending node $\Omega$, argument of
pericenter $\omega$, and true anomaly $\lambda$ were fixed at those of the
asteroid (1644) Rafita at J2000. Synthetic proper elements and frequencies
were computed with the approach described in \citet{carruba_2010}.

\begin{figure}[!htp]
   \centering\includegraphics[width=0.8\linewidth]{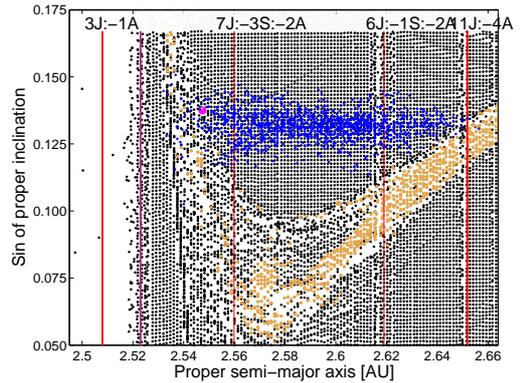}
   \caption{An ($a$,$\sin (i)$) proper element map of the Rafita region.
     Proper elements of test particles are shown as black full dots. The
     $g+g_{5}-2g_{6}$ secular resonance is marked by amber full points. Other
     symbols are the same as in Fig.~\ref{Fig: Orbital_location}.}
   \label{Fig: Dyn_Map}
\end{figure}

Fig. \ref{Fig: Dyn_Map} shows the dynamical map in the proper
($a$,$\sin{(i)}$) plane for 6205 particles in the orbital
neighborhood of Rafita. Black full dots identify the values
of proper elements. Vertical alignments in the map are associated
with mean-motion resonances. Due to their dynamical importance, we choose
to display as light red lines the orbital location of only two resonances
(the 7J:-3S:-2A and 6J:-1S:-2A) listed in \citet{nesvorny_1998}. The proper
elements of real objects in the Rafita HCM dynamical group, obtained after
integrating these bodies with the same scheme used to obtain the dynamical
map, are also plotted over the dynamical map as blue full dots.  Many
resonances cut through the group, and, as a consequence, several family
members are expected to be in these resonances.  A detailed analysis of the
local dynamics was performed in \citet{carruba_2007}, interested readers
could find more details about this region in that paper. Here we just
emphasize the role of the $g+g_{5}-2g_{6}$ secular resonance, which is the
resonance affecting the largest number of asteroids in the region. Objects
whose pericenter frequency is within $\pm 0.3$ arcsec/yr from
$g = 2g_6-g_5 = 52.229$ arcsec/yr, i.e, objects that are more likely to be
in librating states of a four order non-linear secular resonance
\citep{carruba_2009}, are marked in amber.

% ============================
\section{Age of the Rafita family}
\label{sec: Age}

In this section, the method of Yarkovsky isolines is first used to
preliminary assess the possible age of the family and to attempt eliminating
possible dynamical interlopers, i.e. objects whose current orbit could not
be explained as having diffused to its current location to within the
maximum estimate of the family age \citep{carruba_2014,carruba_2015}.
For this purpose, the isolines of equal displacement induced by the Yarkovsky
force were computed for different estimated ages at the position of the
family center. In fact, one problem with this method is what is considered
for the family center. In this work, we computed isolines of maximum
displacement in $a$ for a fictitious family originating at the current
location of 1644 Rafita itself. We used the parameters affecting the
Yarkovsky force for S-type asteroids, as listed in \citet{broz_2013}:
$\rho_{\text{surf}}=1500$ and $\rho_{\text{bulk}}=2000 ~kg~m^{-3}$ for the surface
and bulk densities, $C = 680~J~kg^{-1}~K^{-1}$ for the thermal capacity,
$A = 0.1$ for the Bond albedo, $\epsilon_{IR} = 0.9$ for the thermal
emissivity, and $K = 0.01 ~ W~m^{-1}~K^{-1}$ for the surface thermal
conductivity (with respect to \citet{broz_2013} we are using a larger
value of $K$ and a lower value of $\rho_{\text{bulk}}$ because most of
Rafita members are few-km size asteroids, characterized by higher
values of $K$ and lower densities \citep{Delbo_2015}).  The uncertainty
  by which these parameters is known is, however, one of the main sources
  of error for asteroid family dating \citep{Masiero_2012}.  We warn
  the reader that other possible choices of $K$ and $\rho_{\text{bulk}}$
  could produce different values of the family age.

\begin{figure}
  \centering
  \includegraphics[width=0.8\linewidth]{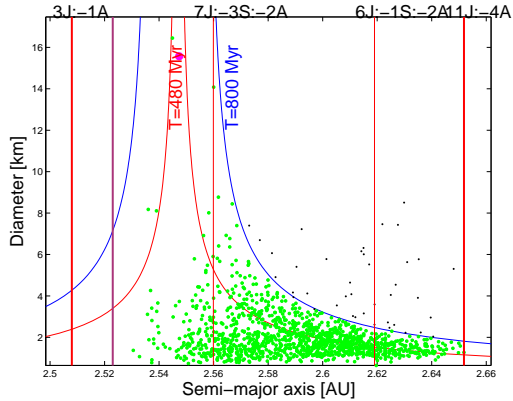}
  \caption{ A proper a versus radius projection of members of the Rafita
    family.  The red and blue lines display the lines of equal Yarkovsky
    displacement for $T=480$ Myr and $800$ Myr, respectively.}
  \label{Fig: yarko_adiam}
\end{figure}

Fig.~\ref{Fig: yarko_adiam} displays the location of the Rafita
family members in the ($a$, $D$) domain. Rafita itself, with a
WISE estimated diameter of 15.57 km, is shown as a magenta full dot.
The concentration of large objects toward the border of the 3:1 mean-motion
resonance indicates that the original structure was affected by this
resonance.  One problem with the application of the method
of isolines of equal displacement caused by the Yarkovsky effect
is what is considered for the family center.  For families
formed in non-catastrophic events, the center can be assumed to
coincide with the largest remnant.  Alternatively, the use
of the center of mass can be used for the cases
of families formed by catastrophic disruption events.

Rafita however is peculiar because i) it was formed in a catastrophic
event, ii) most of the objects on the left side of the family were lost
in the 3J:-1A mean-motion resonance, and iii) (1644) Rafita itself
  could have migrated by up to 0.005 au outward or inward over the
  estimated $\simeq 0.5$~Gyr age of the family, assuming a maximum
  Yarkovsky drift rate for an asteroid of this size, and $0^{\circ}$ or
  $180^{\circ}$ stable spin obliquity.  Since the current barycenter
  of the Rafita family is not a viable option because of the family
  incompleteness, we computed isolines of maximum displacement in $a$
  for a fictitious family originating at the current location of 1644
  Rafita itself (Fig. \ref{Fig: yarko_adiam}).  Errors possibly
  caused by the Yarkovsky induced uncertainty in the posotion of
  1644 Rafita will be accounted for later on in this paper.

Isolines were computed for times equal to
480 Myr, the estimated age according to \citet{nesvorny_2015,carruba_2016a}
using a V-shape criterion, and 800 Myr respectively, which is
quite larger than the maximum estimate of the family age (580 Myr) found
in the literature.  Since this method does not account for the
initial dispersion of the family members, however, ages obtained with
this method can be overestimated.  The 43 asteroids outside the maximum
Yarkovsky isoline in Fig. \ref{Fig: yarko_adiam} were considered
dynamical interlopers and excluded from the family list.

\begin{figure*}
  \centering
  \begin{minipage}[c]{0.49\textwidth}
    \centering \includegraphics[width=3.1in]{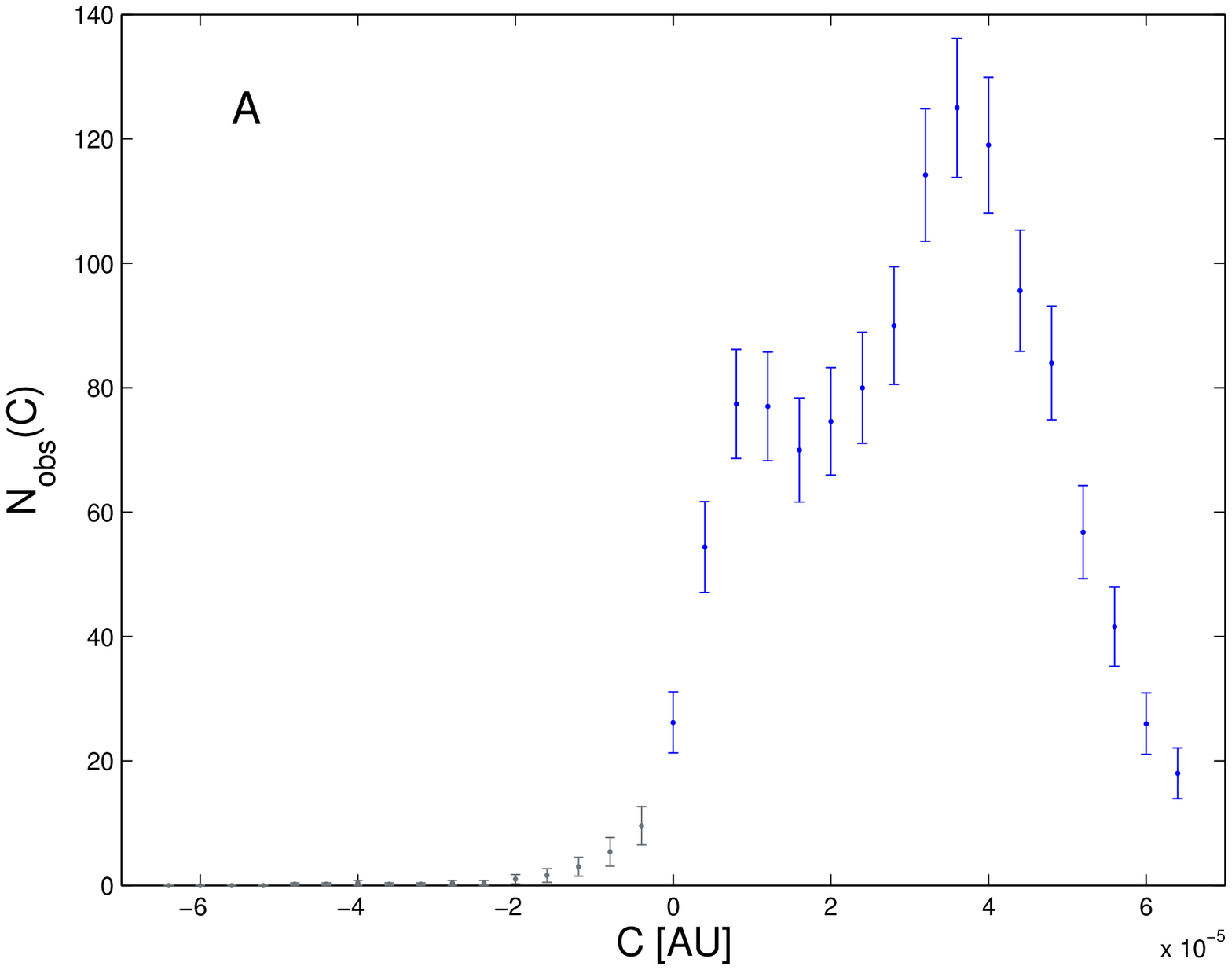}
  \end{minipage}%
  \begin{minipage}[c]{0.49\textwidth}
    \centering \includegraphics[width=3.1in]{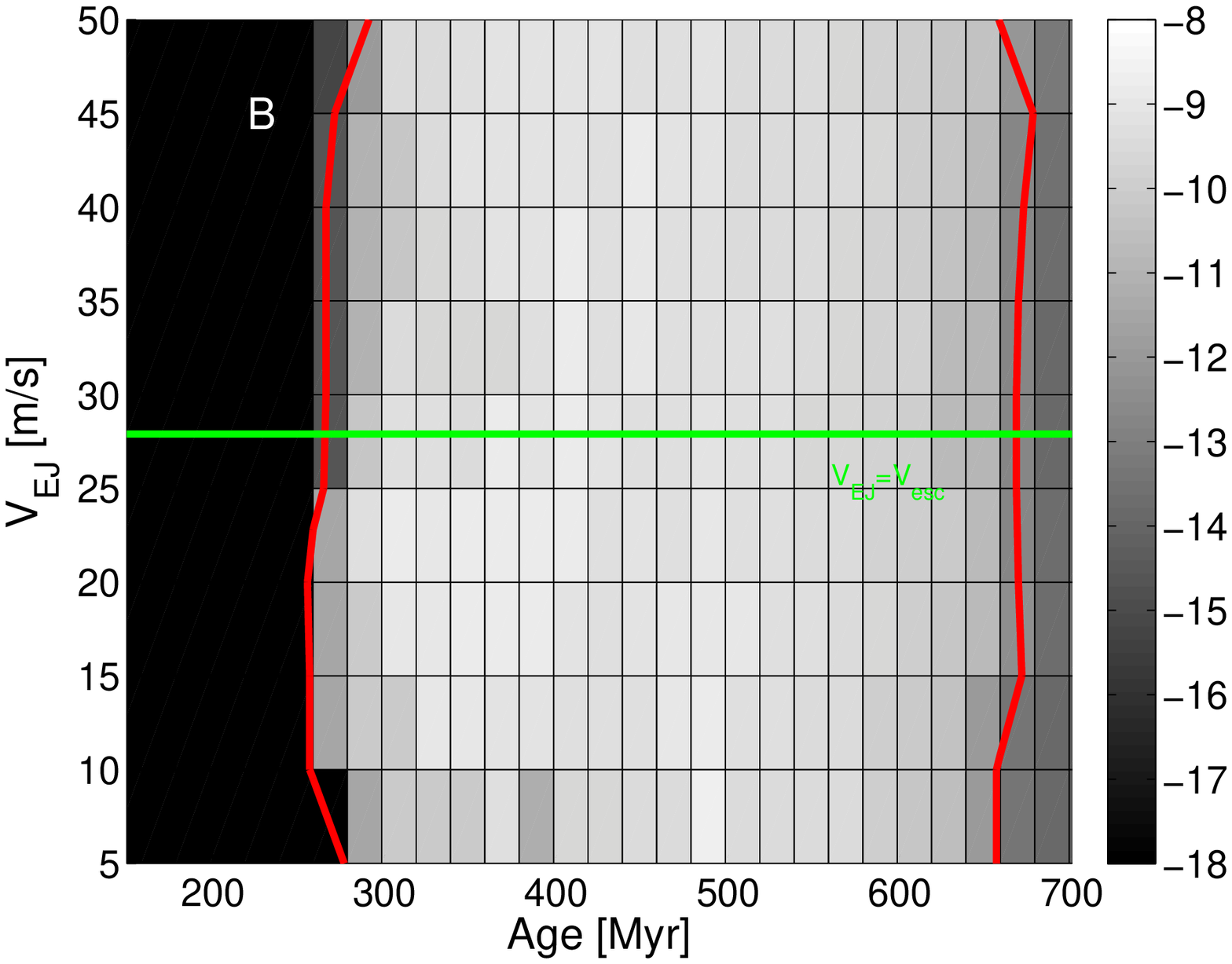}
  \end{minipage}
   \caption{Panel A: a histogram of values of the C target function from
     Eq.~\ref{eq: C} for the Rafita family.  Panel B: Values of the
     $\psi \Delta C$ function in the (Age, $V$)plane, for simulated Rafita
     families.}
   \label{Fig06_target_function_C}
\end{figure*}

Yarkovsky isolines do not provide an optimal estimate of asteroid families
ages, since they neglect the effect of the initial ejection velocity field.
However, they can be used to obtain a preliminary value that can be later
refined by a more advanced method.  In the present work we also used the
Yarko-Yorp Monte Carlo method of \citet{vokrouhlicky_2006a,vokrouhlicky_2006b,
  vokrouhlicky_2006c}, to obtain a more precise estimation of
the age and ejection velocity parameter.  We modeled the evolution of
simulated family members due to Yarkovsky and YORP forces over time intervals
of the order of the age of the family. This method was modified by
\citet{carruba_2015} to include the stochastic version of the YORP effect
\citep{bottke_2015}. Modeling the diffusion via YORP effects, a target
function $C$ for each body is defined as:

\begin{eqnarray}
  0.2H=\text{log}_{10}(\frac{\Delta a}{C})
  \label{eq: C}
\end{eqnarray}

\noindent where $H$ is the absolute magnitude of the body
and $\Delta a = a-a_{b}$, where $a_{b}$ is the family center, assumed
equal to the current position of 1644 Rafita. The distribution of the
family members can be characterized using a one-dimensional array
$N_{obs}(C)$ indicating the number of asteroids in the interval
$(C, C +\Delta C)$. The histogram of the target function C for Rafita family
presented as blue line in Fig.~\ref{Fig06_target_function_C}, panel A.  33
intervals are used starting at $C_{\text{min}} = -6.6 \times 10^{-5}$
with a step of $4.0\times 10^{-6}$ au.  Errors are assumed to be proportional
to the square root of the number of asteroids in each $C$ bin.  To account
  for the uncertainty in the original orbital position of Rafita, we
  computed an average of five $C$ distribution, one centered at the current
  orbital position of Rafita, and 2 each at $\pm 0.5$ and $\pm 1$ of
  the maximum drift in $a$ caused by the Yarkovsky effect (0.005 au).
Because the Rafita family is currently incomplete at lower values of
semi-major axis
(and therefore of $C$), we neglected negative $C$ values in our analysis
(shown as gray in \ref{Fig06_target_function_C}).  Since we are only using
the positive $C$ distribution to estimate two parameters, our distribution
has 15 (17-2) degrees of freedom.

Following the work of \citet{carruba_2014, carruba_2015, carruba_2016b},
fictitious distributions of asteroids were evolved for different values of
the ejection velocity parameter $V_{EJ}$ (see for instance Eq.~3
in \citet{carruba_2016b} for a definition of this parameter), considering
the Yarkovsky effect (both diurnal and seasonal versions), and the
stochastic YORP torque.  Assuming that about half of the family was lost in
the 3J:-1A mean motion resonance, and considering a density of S-type bodies,
a parent body of 45 $km$ in diameter was considered to estimate an escape
velocity as $V_{esc}=28 ~m.s^{-1}$. During the investigation of the shape of
the ejection velocity field of 49 asteroid families by \citep{carruba_2016a},
it was found that the typically observed values of $\beta=\frac{V_{EJ}}{V_{esc}}$
did not generally exceed 1.5, Thus, the maximum values of $V_{EJ}$ here
considered will be 50 $\text{m}.\text{s}^{-1}$.  Results for
$V_{EJ} > 50 \text{m}.\text{s}^{-1}$ will not be shown, for simplicity.
The distributions of the C function for the simulated family is then
compared with that for the real one (supposing a half of it was lost
by the 3J:-1A mean motion resonance), minimizing a $\chi^{2}$-like
function $\psi \Delta C$ (see Eq.~6 in \citet{carruba_2016b})

The values of $\psi \Delta C$ in the (Age, $V$) plane are presented in
Fig. \ref{Fig06_target_function_C}, panel B.  For fifteen degrees of 
freedom, and assuming that the $\psi \Delta C$ follows an
incomplete gamma function distribution \citep{press_2001}, 
the value $\psi \Delta C = 13.53$ (red line in Fig. 
\ref{Fig06_target_function_C}, panel B) is associated with a one sigma 
probability of 68.3\% that the simulated family and the real one 
are compatible.  Our results are compatible with the estimate
from \citet{nesvorny_2015}:  we find an age of $490^{+190}_{-220}$ Myr,
and a value of the ejection velocity parameter of $20^{+50}_{-20}$~m/s,
compatible with the estimated escape velocity from the parent body
($\beta = 0.71$).  In the next section we will analyze the 
dynamical evolution of fictitious families obtained with
this value of $V_EJ$, and discuss what constraints on the
family age can be obtained by these numerical experiments.

\section{Dynamical evolution of the Rafita family} 
\label{Dynamical_evolution}

Numerical integration with the \textsc{swift\_rmvsy} code from the SWIFT
package \textsc{swift} \citep{levison_1994}, that uses the Regularized Mixed
Variable Symplectic method, modified by \citep{broz_1999}, were performed
over 800 Myr in order to investigate the dynamical evolution of Rafita
fictitious family members, obtained with the approach described in the 
previous section and $V_{EJ} = 20$~m/s. The \textsc{swift\_rmvsy} 
integrator allows to integrate a set of massless test
particles (that do not interact among themselves) under the gravitational
influence of all planets, considering the diurnal and seasonal versions of
the Yarkovsky effect.  Using the method outlined in \citet{machuca_2012}, 
two spin axis obliquities forming angles of $0^{\circ}$ and $180^{\circ}$
  with the orbital angular momentum were used to investigate the maximum
orbital mobility
caused by the Yarkovsky effect.  Synthetic proper elements for each test 
particles were obtained with the approach described in \citet{carruba_2010}. 

\begin{figure*}
    \centering
  \begin{minipage}[c]{0.49\textwidth}
    \centering \includegraphics[width=3.1in]{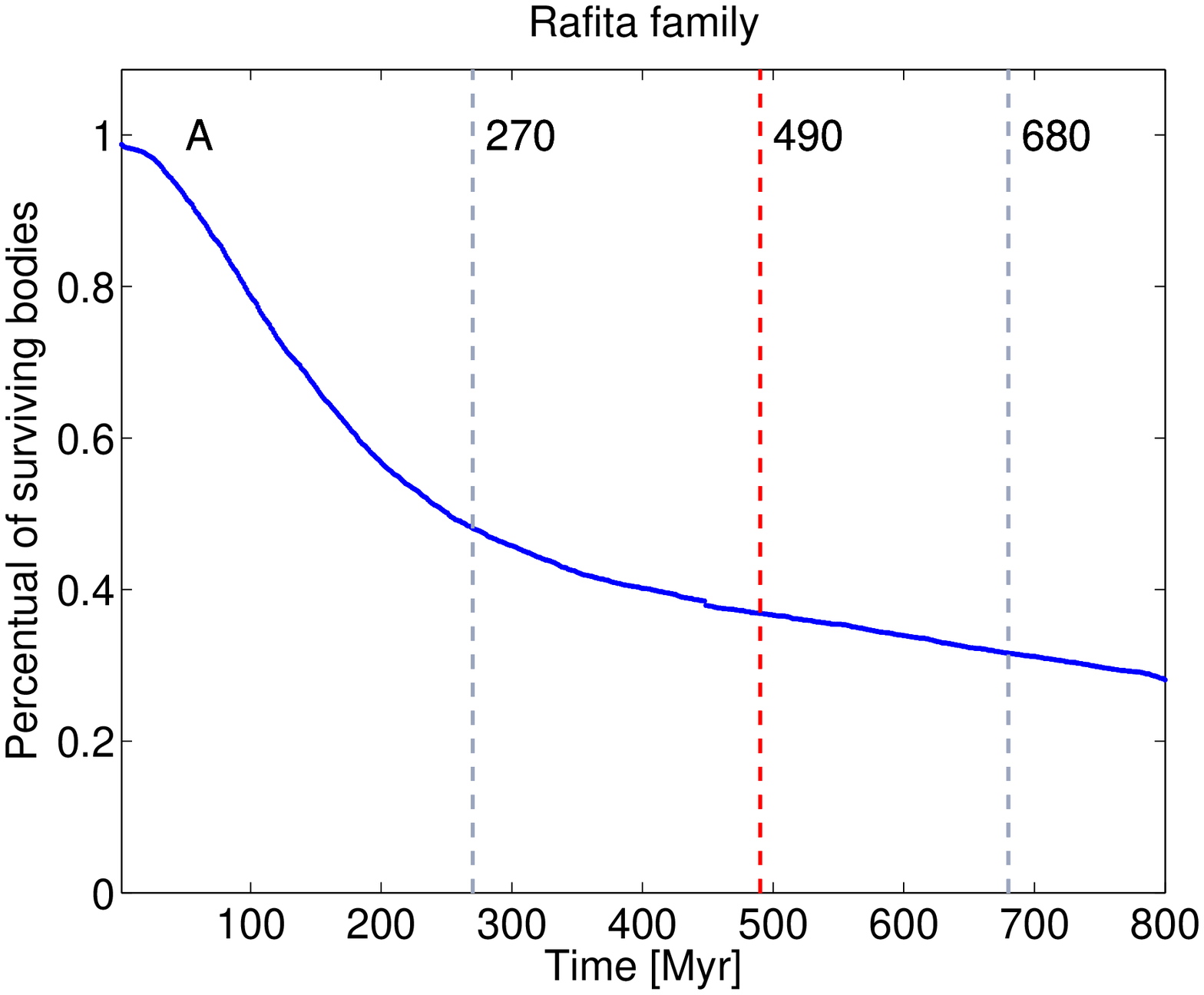}
  \end{minipage}%
  \begin{minipage}[c]{0.49\textwidth}
    \centering \includegraphics[width=3.1in]{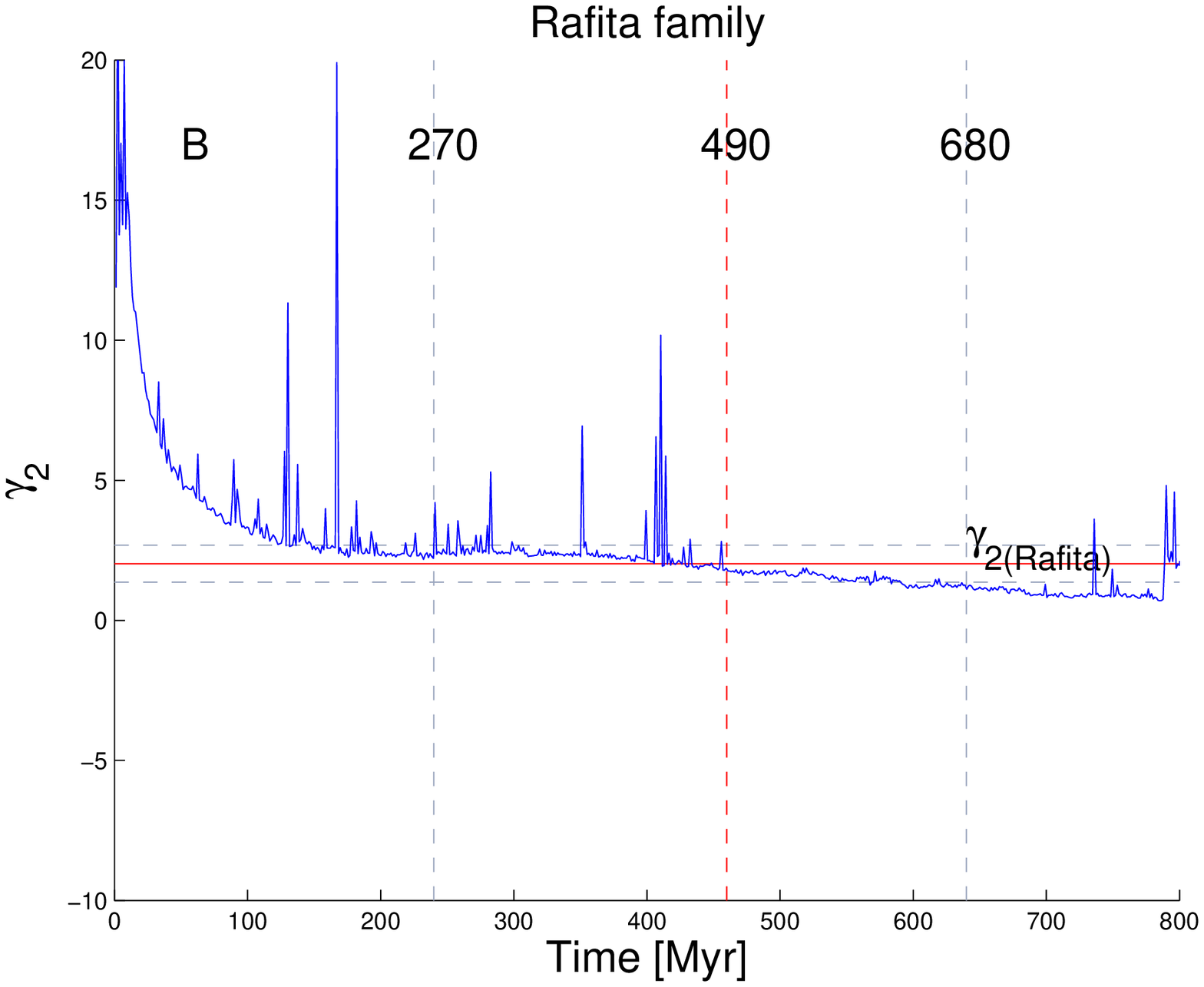}
  \end{minipage}

  \caption{Panel A: percentile of surviving particles in the Rafita region
    as a function of time.
Vertical dashed lines display the estimated age of the family, according
to this work.  Panel B: time behavior of the kurtosis of the 
$v_W$ component of the ejection velocity field (${\gamma}_2{(v_W)}$) for 
the simulated Rafita family.  The horizontal red line display the 
current value of the ${\gamma}_2{(v_W)}$ parameter for the Rafita
group, while the horizontal dashed lines show the current value plus
or minus its error.}
\label{Fig08_N_part_tim}
\end{figure*}

Defining a Rafita orbital region with the method described in 
Sect.~\ref{sec: fam_ide}, we first checked the number
of particles that remained in the region as a function of time.
Results are shown in Fig.~\ref{Fig08_N_part_tim}, panel A, where we display
the percentile of surviving particles in the Rafita region as a function
of time.  The vertical red dashed line is associated with the estimated
family age of of 490 Myr, while the other vertical dashed lines are
associated with the minimum and maximum estimated family age, according
to this work.  At the estimated family age, only 36.9\% of 
the original family members remained in the Rafita region, which 
indicates that more than half of the family was lost because
of dynamical mechanism, mainly interaction with the 3J:-1A mean-motion
resonance.  Since recently \citet{carruba_2016c} introduced
a method based on the time behaviour of the kurtosis of the 
$v_W$ component of the ejection velocity field (${\gamma}_2{(v_W)}$
of families characterized by a leptokurtic distribution of this
parameter, and since the Rafita family is one of the most 
leptokurtic family in the main belt (${\gamma}_2{(v_W)}$ = 0.72, 
\citet{carruba_2016a}), we also checked the time behavior of this 
parameter for the simulated family.  Results are shown in 
Fig.~\ref{Fig08_N_part_tim}, panel B. Errors for the ${\gamma}_2{(v_W)}$
  were computed assuming that they were equal to
  $\frac{\sqrt{24 {{\sigma}(v_W)}^2}}{n}$, where ${\sigma}(v_W)$ is the standard
deviation in $v_W$ and $n$ is the number of Rafita members
\citet{Kendall_1969}.
Except for isolated spikes, 
associated with single particles that experienced sudden changes of 
inclination, the current value of ${\gamma}_2{(v_W)}$ was reached in 
timescales compatible with those associated with the family estimated age.
\footnote{Results for this simulations were obtained under the assumption
  that the asteroid spin obliquity remain fixed at $0^{\circ}$ or $180^{\circ}$
  during the simulation.  Including YORP random walking may affect the
  dispersion in inclination, and therefore affect the observed values
  of ${\gamma}_2{(v_W)}$.  While we do not expect this effect to dramatically
  alter the time evolution of the ${\gamma}_2{(v_W)}$ on short timescales,
  this may be important (or not) at longer timescales.  Assessing the
  importance of this effect for more evolved asteroid families remain,
  in our opinion, a challenge for future work.}.
  
\begin{figure}
  \centering
  \includegraphics[width=1\linewidth]{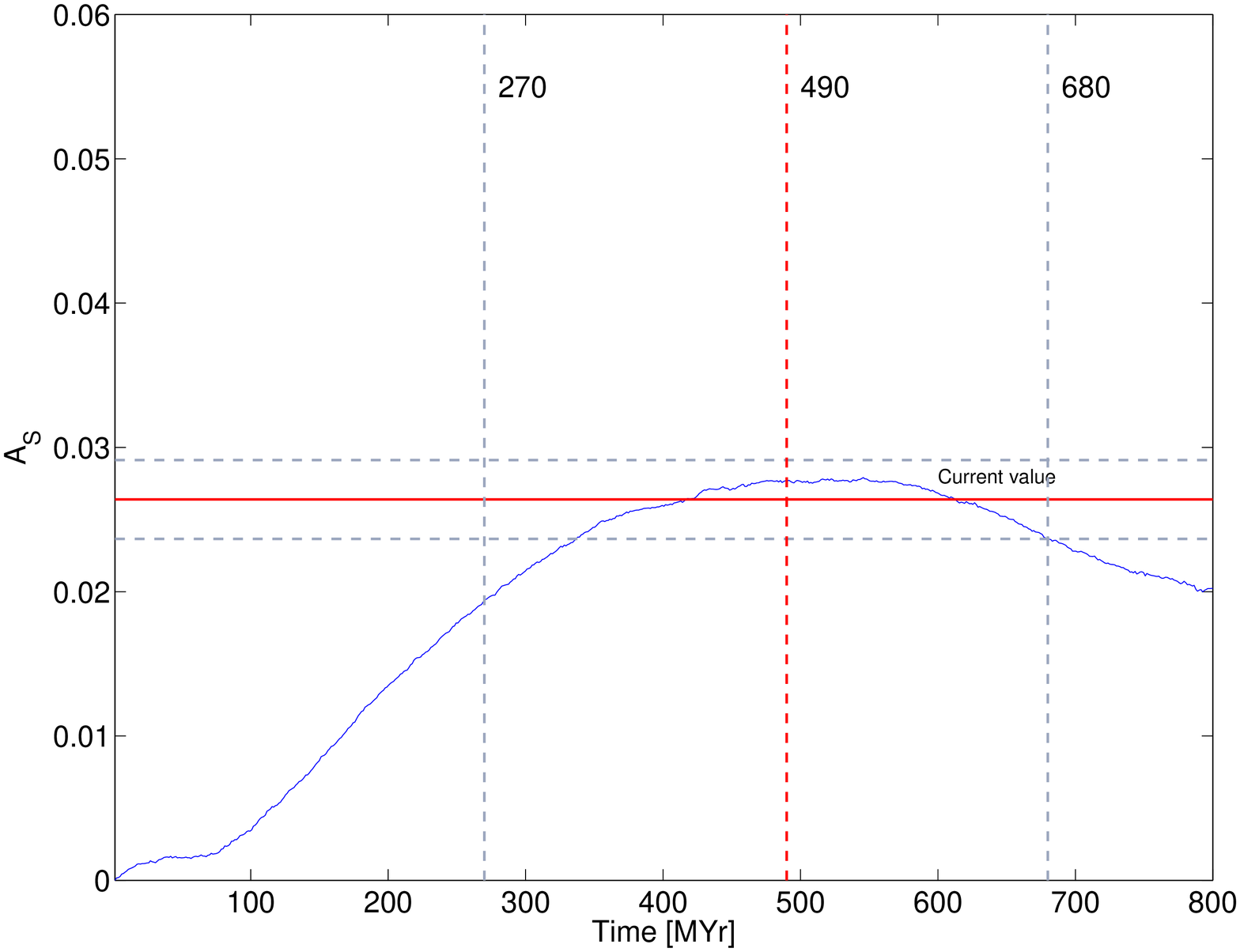}
  \caption{ Asymmetry coefficient of the target function $C$,
for the case where the center of the family was assumed to be
at the current location of (1644) Rafita. 
The horizontal lines display the current value of $A_S$ for the Rafita family.
Dashed horizontal lines display the current value of $A_S$ plus 
or minus its error, obtained with standard error propagation formulas.
Vertical lines have the same meaning as in Fig.~\ref{Fig08_N_part_tim}.}
\label{Fig09_sym_coef}
\end{figure}

Motivated by this preliminary analysis, we continued our analysis by 
analyzing the degree of asymmetry of the current Rafita family.
As observed in Fig. \ref{Fig06_target_function_C}, panel A, the current
distribution of $C$ values for the Rafita family is considerably
asymmetric, with much fewer asteroids in negative $C$ value bins 
than in positive ones.  To quantify the degree of asymmetry 
of the Rafita family, we introduce an asymmetry coefficient $A_S$ defined
as:

\begin{equation}
A_S=\frac{1}{N_{ast} \cdot N_{int}}\sum_{1}^{N_{int}}(N_{pos}(i)-N_{neg}(i)),
\label{eq: asym_coeff}
\end{equation}

\noindent

where $N_{ast}$ is the number of objects in the simulated
Rafita family, $N_{int} = 17$ is the number of positive interval in
the $C$ distribution,
and $N_{pos}(i)$ and $N_{neg}(i)$ are the number of asteroids in the positive
and negative $i-th$ bin of the distribution.  Family with a larger
population in positive $C$ bins would have a positive $A_S$ value,
while the opposite would be true for families with larger populations
in negative $C$ bins.  The current values of $A_S$ for the Rafita family,
when its center is assumed to be at the current orbital location of
(1644) Rafita, is 0.026$\pm$0.003. We computed values of $A_S$ for our
simulated family using the current orbital location of (1644) Rafit
as a reference as a function of time.  Our results are shown in
Fig.~\ref{Fig09_sym_coef}.  Remarkably, we find a very good
agreement with the age estimated from the
analysis of $A_S$ and that obtained from the other two previously described
methods.  In the next section we will estimate the contribution
to the NEA population from the Rafita family.

\section{The contribution to the NEA population from the Rafita family}
\label{sec: NEA_Rafita}

Following the techniques described in \citet{masiero_2015a}, we have
performed numerical simulations of the evolution of small family
members from the Rafita breakup event into near-Earth space.  Family
members were randomly generated with initial ejection velocities
matching our best-fit velocity of 20 m/s around the current orbit of
the presumed parent body (1644) Rafita.  Particles were given the physical
in the previous sections, a geometric albedo following
the family mean of $p_V$= 0.26 and a diameter randomly drawn from the
current family size frequency distribution (\citet{masiero_2015b}),
between 1 and 7 km. 

Family members were forward-integrated for 500 million years using
\textsc{swift\_rmvsy}.  Spin axes were initially randomized, and
evolved under YORP and collisional reorientation.  When a family
member entered near-Earth space ($q<1.3$ AU) its instantaneous orbit
every 10 kyr was recorded until it left near-Earth space or was
removed from the simulation (either via collision with the Sun or a
planet, or through ejection from the Solar system).  These orbit
snapshots were compiled to build a probability map of where NEOs
originating from this family would be found in orbital element-space.

\begin{figure*}
    \centering
  \begin{minipage}[c]{0.49\textwidth}
    \centering \includegraphics[width=3.1in]{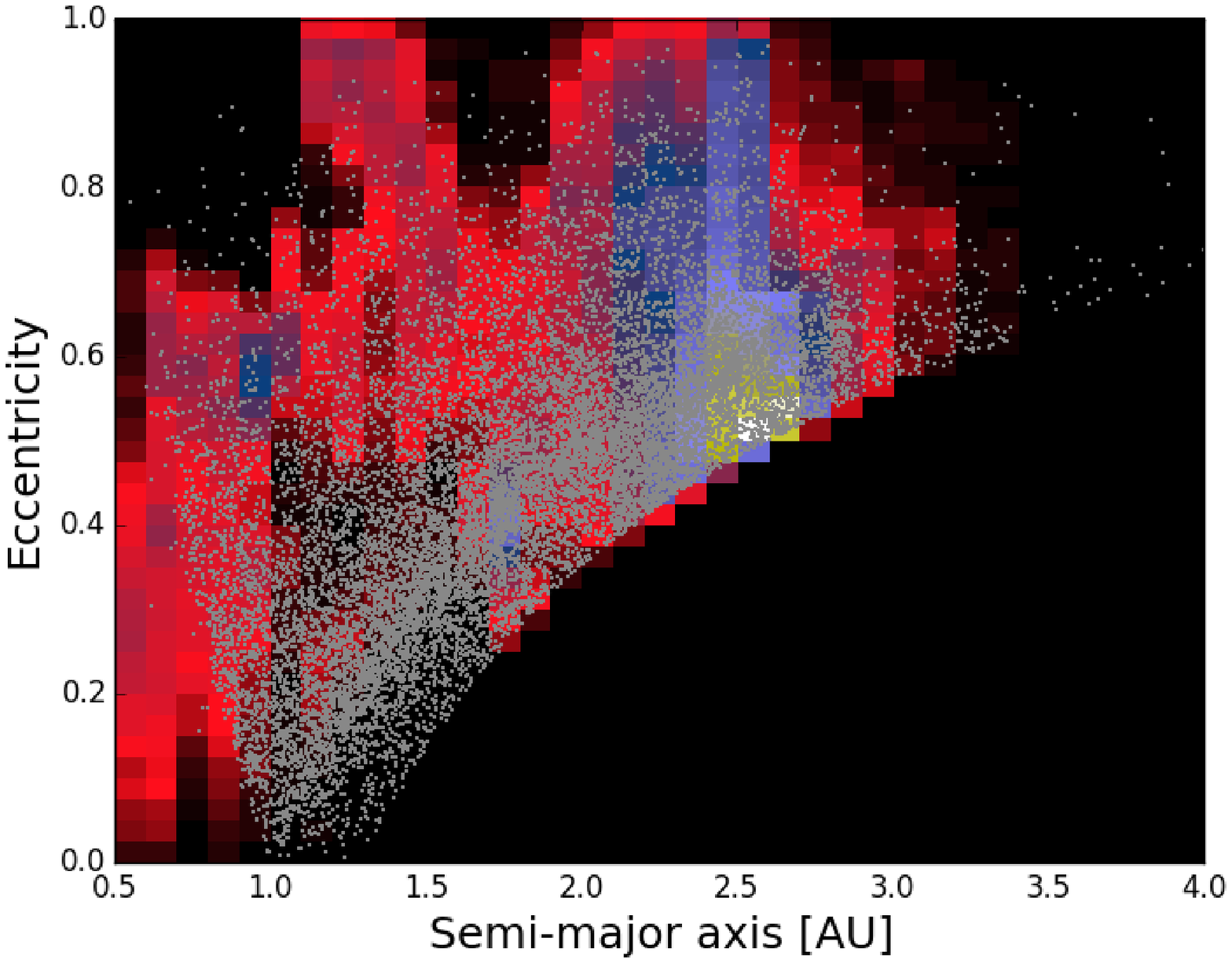}
  \end{minipage}%
  \begin{minipage}[c]{0.49\textwidth}
    \centering \includegraphics[width=3.1in]{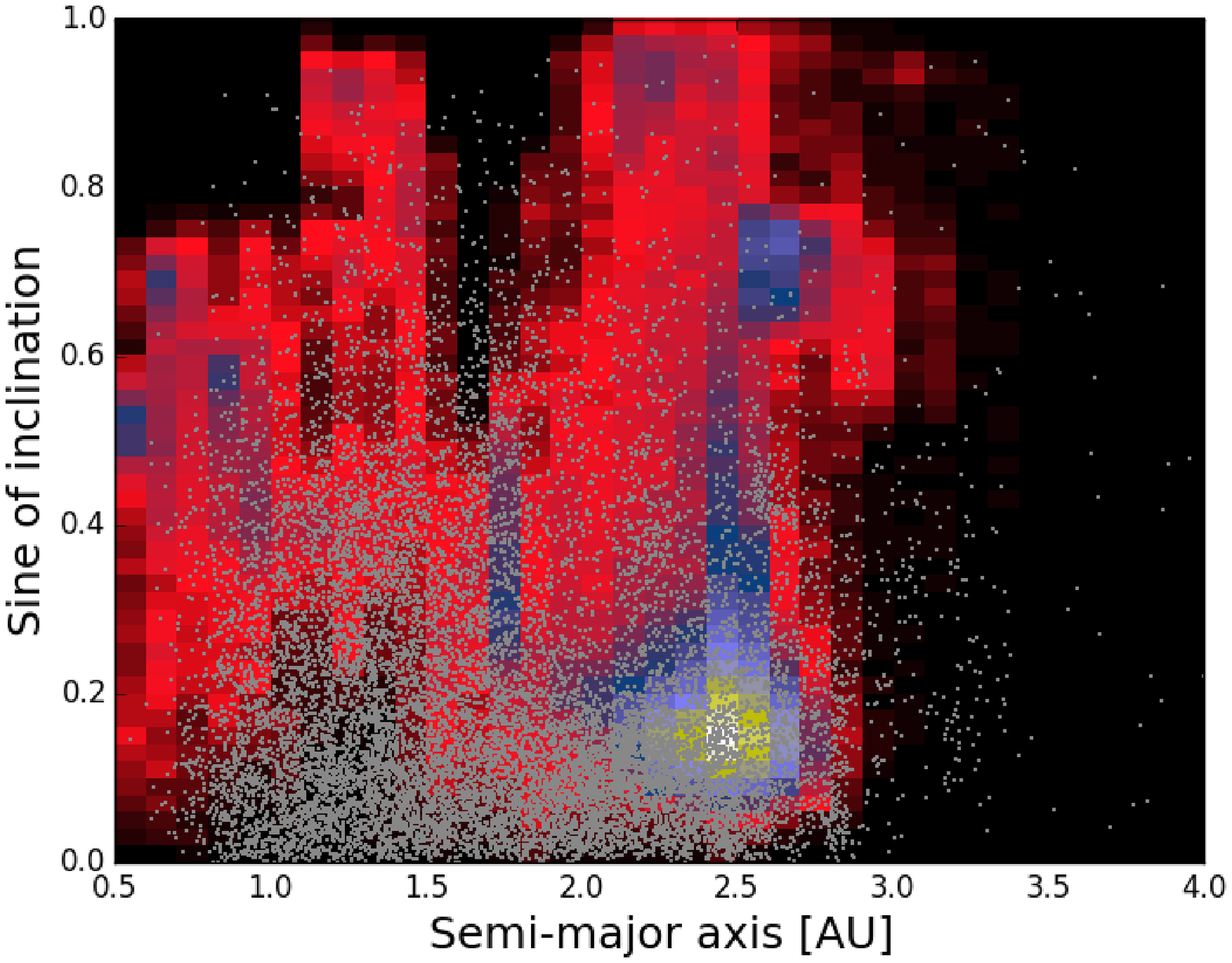}
  \end{minipage}

  \caption{Probability map of near Earth asteroids originating from
      the Rafita family (background heat map) for semimajor axis versus
      eccentricity and semimajor axis versus the sine of inclination.
      Grey points indicate the orbits of all currently known NEAs.}
\label{fig: Masiero}
\end{figure*}

We show in Fig.~\ref{fig: Masiero} the preliminary probability maps from a test
simulation.  The highest probability spatial bins in each plot are shown in
white (normalized probability of 1), with yellow, purple (50\%), blue and
finally red showing decreasing probability density.  Bins shown in black had
zero or near-zero recorded contribution from the simulated family members.
 Simulations of NEO evolution were conducted spanning a range of probable
  bulk densities (1500-2500 $kg~m^-3$), but the simulated density had no
  significant effect on the resulting probability density map.  The only
  difference of note is that due to the lower densities, the smallest
  objects in the simulation were more quickly moved into NEO-space, thus
  the overall flux of asteroids was a factor of $\simeq$4 higher early in the
  simulations, decreasing to $\simeq$1.5 times higher by the current age of
  the family as the slower, high density test objects finally reach the
  resonances that push them into NEO space.

Future works will use a larger number of simulations to determine probability
densities with higher accuracy,
as individual runs can be dominated by a single long-lived object.
In general, we find that NEAs from Rafita enter near-Earth space at similar
semi-major axes and inclinations as the family ($a\sim2.5$,
$i \sim 8^{\circ}$) as a result of eccentricity pumping.
This phase space overlaps 8\% (1272 out of 15091) of presently known NEAs.
There is an immediate surge near $T = 0$~Myr in the NEAs population
  from the small objects that are placed into the nearby resonances by
  the assumed impact generated velocities.  After that, however, the family
  follows a steady injection rate for the rest of the simulation.
Further followup would be, however, required to confirm a direct link between
the NEOs and the Rafita family, since alternative pathways to
produce these objects should also be studied.  Quite interestingly, 4\%
of the simulated objects that survived the last 30 Myr of the simulation
seem to have rather different orbits than the general heat map shows,
with a cluster near 1~AU.  Confirming or denying the potential contribution
of Rafita members to the observed population of Potentially Hazardous
Asteroids (PHA) remains an interesting challenge for future works.

\section{Conclusion}
\label{sec: concl}

Our results could be summarized as follows:

\begin{itemize}

\item Revised the current knowledge on the Rafita asteroid family.
This group is an S-type family, with few possible C-complex interlopers,
and was most likely the product of a catastrophic disruption of a 
$D \simeq 45$ km parent body. The Rafita family is cut on the 
left side in $a$ by the 3J:-1A mean-motion resonance, and interacts
with the $g+g_5-2g_6$ secular resonance.

\item We used the method of Yarkovsky isolines to eliminate possible
dynamical interlopers, and Monte Carlo methods simulating the 
dynamical evolution of several fictitious asteroid families
under the influence of the Yarkovsky and YORP effects.  We find an 
age of $490^{+190}_{-220}$ Myr, and a value of the ejection velocity
  parameter $V_{EJ}$ of $20^{+50}_{-20}$~m/s, comparable with the estimated
escape velocity from the parent body ($\beta = \frac{V_{EJ}}{V_{esc}} = 0.71 $).

\item We studied the dynamical evolution of fictitious members
of the Rafita asteroid family simulating the initial conditions
after break-up of the parent body and obtained new estimates
of the family age using the method of the time dependence
of the ${\gamma}_2{(v_W)}$ parameter \citep{carruba_2016c}
and a new method based on the time dependence of the coefficient
$A_S$ describing the asymmetry of the $C$ distribution of
the simulated Rafita family. Remarkably, these two methods
provide estimates in good agreement with the results of
the Yarko-Yorp approach.

\item We studied the possible contribution to the NEA population
  from the Rafita family.  1\% of the simulated particles are NEOs in
  the last 10 Myr of the simulation, and $\simeq$40\% of these cluster
  with semi major axes near 1 AU and could potentially contribute to the
  current population of Potentially Hazardous Asteriods (PHA).

\end{itemize}

Overall, our results confirm and refine previous estimates of the 
Rafita family age, with a new estimate of $490\pm 200$ Myr.
More important, we introduced a new method to date the age of incomplete
asteroid families based on the asymmetry coefficient of their $C$ function 
distribution, that provided estimates of the family age in agreement
with independent method based on Monte Carlo simulations of the
Yarkovsky and YORP dynamical evolution of family members and
of the time evolution of the ${\gamma}_2{(v_W)}$ parameter.
This new method can be potentially applied to other incomplete asteroid
families, such as the case of the Maria (interaction with the 3J:-1A 
mean-motion resonance) and Gefion (interaction with the 5J:-2A 
mean-motion resonance) families, and, in our opinion, represent the main
result of this work.

% ============================
\section*{Acknowledgments} 

We are grateful to the reviewer of this paper, Dr. David Vokrouhlick\'{y},
for comments and suggestions that greatly improved the quality of this paper.
The authors wish to thank the S\~ao Paulo State Science Foundation (FAPESP)
(Grants 13/15357-1 and 14/06762-2) and the Brazilian National Research
Council (CNPq, grant 305453/2011-4) for the generous support of this work.
JRM was funded through the JPL internal Research and Technology Development
program.
This publication makes use of data products from the Wide-field 
Infrared Survey Explorer (WISE) and NEOWISE, which are a joint project 
of the University of California, Los Angeles, and the Jet Propulsion 
Laboratory/California Institute of Technology, funded by the National 
Aeronautics and Space Administration.

% \appendix
% \section[]{xxx}

\label{lastpage}

\end{document}